\documentclass[prl,a4paper]{revtex4}

\usepackage{amssymb}
\usepackage{graphicx}
\usepackage{amsmath}
\usepackage{amsbsy}
\usepackage{amsthm}
\usepackage[ansinew]{inputenc} %umlaute
\usepackage{bbm}
\usepackage{bm}
\usepackage{epsfig}
\usepackage{lscape}
\usepackage{geometry}
\newcommand{\id}{\mathbbm{1}}

\newcommand{\beq}{\begin{eqnarray}}
\newcommand{\eeq}{\end{eqnarray}}
\newcommand{\bra}[1]{\ensuremath{\langle #1 |}}
\newcommand{\ket}[1]{\ensuremath{| #1 \rangle}}

\begin{document}
\title{Device-Independent Test for Genuine Multipartite Entanglement}
\author{Andreas Gabriel$^1$}
%\email{andreas.gabriel@univie.ac.at}
\author{{\L}ukasz Rudnicki$^2$}
\author{Beatrix C. Hiesmayr$^{1}$}
\email{Beatrix.Hiesmayr@univie.ac.at}
\affiliation{$^1$ University of Vienna, Faculty of Physics, Boltzmanngasse 5, 1090 Vienna, Austria\\
$^2$ Center for Theoretical Physics, Polish Academy of Sciences, Aleja Lotników 32/46, PL-02-668 Warsaw, Poland}

\begin{abstract}
We investigate a-priori detection probabilities of genuine multipartite entanglement (GME). Even if one does not have knowledge about the basis in which a state is produced by a source, how a channel decoheres it or about the very working of the detectors used, we find that it is possible to detect GME with reasonably high probability in a feasible fashion. We show that by means of certain separability criteria, GME can be detected in a measurement-device-independent way. Our method provides several applications whenever e.g. state tomography is not possible or too demanding, and is a tool to investigate security issues in multi-particle quantum cryptographical protocols.
\end{abstract}

\maketitle

\section{Introduction}
Entanglement is one of the most intriguing and most fundamentally nonclassical phenomena in quantum physics. One important type of entanglement - both for research concerning the foundations of quantum theory (see e.g. \cite{foundations1,foundations2}) and for conceivable technological applications such as quantum computers \cite{computers,computers2,computers3} or quantum cryptography schemes \cite{crypto1,qss} - is genuine multipartite entanglement (GME). It is even assumed that entanglement (and, particularly, GME) might play significant roles in Nature on all scales, ranging from (comparatively) elementary effects like phase transitions \cite{gmephasetrans} or frustration \cite{frustration} in crystals to complex mechanisms in lifeforms, e.g. photosynthesis \cite{gmephotosynth} or even geographical orientation of birds \cite{birds}. In more and more physical systems multipartite and high dimensional entanglement is observed, e.g. in neutron interferometry~\cite{HiesmayrGMENeutron} or for spatially entangled photons (e.g. Refs.~\cite{Loeffler1,Loeffler2}).

One of the major tasks in entanglement theory is entanglement detection. Given a quantum state, there are many ways to detect its different entanglement properties - ranging from generic entanglement detection (see e.g. \cite{entdet1,entdet2,entdet3,entdet4,entdet5}) to more specific forms of entanglement, such as GME (see e.g. \cite{gmedet1,gmedet2,gmedet3}). However, it is much more difficult to make any statement about entanglement in a not (entirely) known state (see e.g. \cite{devindep1}). The aim of this work is to present a method for detecting GME in partially unknown states which can be interpreted as states from a partially unknown source, or as measuring the states with untrusted measurement equipment.

This work is organised as follows. In the upcoming section, basic definitions on GME will be reviewed and the separability criteria on which our scheme is based are defined. After that, we present our method of device-independent GME detection, which is the main result of this paper and is thoroughly discussed in the following sections, before the paper is concluded.

\begin{figure}[ht!]\centering\includegraphics[width=0.55\textwidth]{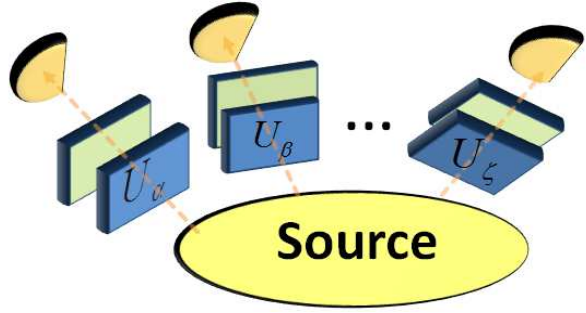}
\caption{Sketch of our device-independent test for a priori detection probabilities of genuine multipartite entanglement.}\label{sketch}\end{figure}

\section{Genuine Multipartite Detection Criteria}

In order to present our results, we need to introduce some basic definitions on multipartite entanglement and its detection first.\\
A pure multipartite quantum state $\ket{\Psi_2}$ is called biseparable, iff it can be written as a nontrivial product: $\ket{\Psi_2} = \ket{\psi_1}\otimes\ket{\psi_2}$, where the $\ket{\psi_i}$ are states of one or several subsystems. A mixed multipartite state $\rho_2$ is called biseparable, iff it can be written as a mixture of biseparable pure states: $\rho_2 = \sum_i p_i \ket{\Psi_2^i}\bra{\Psi_2^i}$, where the $\{p_i\}$ form a probability distribution (i.e. $p_i \geq 0$ and $\sum_i p_i=1$) and the $\ket{\Psi_2^i}$ may be biseparable under different bipartitions (consequently, the mixed biseparable state is not necessarily separable under a specific bipartition). Any state that is not biseparable is called genuinely multipartite entangled (GME).\\
Detecting GME in $n$-partite mixed states can be a rather challenging task, as both biseparable and GME states can be inseparable under all bipartitions and thus hard to distinguish. One way of approaching this problem is by introducing convex detection criteria, which effectively reduce the mixed-state-problem to the more simple pure-state-problem. A set of such criteria was introduced in Refs.~\cite{hmgh,huber_dicke} via a set of quantities $Q_i$ ($0 \leq i \leq \left\lfloor\frac{n}{2}\right\rfloor$).

The first criterion for a $n$-partite state is given by
\beq Q_0(\rho)=|\bra{0}^{\otimes n}\rho\ket{1}^{\otimes n}| - \sum_{\gamma}\sqrt{\bra{0}^{\otimes n}\bra{1}^{\otimes n}\mathcal{P}_{\gamma_A}^\dagger \rho^{\otimes 2} \mathcal{P}_{\gamma_A} \ket{0}^{\otimes n}\ket{1}^{\otimes n}} \eeq
where the sum runs over all bipartitions $\gamma = \{A, B\}$. The permutation operators $\mathcal{P}_{\gamma_A}$ permute the two copies of all subsystems contained in the first part of $\gamma$.

A set of other criteria is denoted by
\beq Q_m(\rho) = \sum_{\sigma}\left(|\bra{d_\alpha}\rho\ket{d_\beta}| - \sqrt{\bra{d_\alpha}\bra{d_\beta} \mathcal{P}_\alpha^\dagger\rho^{\otimes 2} \mathcal{P}_\alpha \ket{d_\alpha}\ket{d_\beta}}\right) - m(n-m-1)\sum_\alpha\bra{d_\alpha}\rho\ket{d_\alpha} \eeq
for $1 \leq m \leq \left\lfloor\frac{n}{2}\right\rfloor$, where the first sum runs over all sets $\sigma = \{\alpha,\beta\}$ with $\alpha,\beta \subset \{1,2,...,n\}$ such that $|\alpha|=|\beta|=m$ and $|\alpha \cap \beta| = (m-1)$, $\ket{d_\alpha}$ is the product state vector with $\ket{1}$ in all subsystems $i \in \alpha$ and $\ket{0}$ otherwise, i.e.
 \begin{eqnarray}
 |d_\alpha\rangle &=& \bigotimes_{i\not\in \alpha} |0\rangle_i\bigotimes_{i\in\alpha} |1\rangle_i\;.
 \end{eqnarray}
The permutation operators $\mathcal{P}_\alpha$ permute all subsystems in $\alpha$ with their respective copies in the two-copies of $\rho$.\\
All these quantities $Q_i$ are by construction nonpositive for biseparable states, i.e. any assumed positive value for any $\rho$ implies this $\rho$ to be GME.

As an example, let us consider the GHZ-state for $n$ qubits in the computational basis \begin{eqnarray}\label{ghz}|GHZ_n\rangle=\frac{1}{\sqrt{2}}\left(|0\rangle^{\otimes n}+|1\rangle^{\otimes n}\right)\;,\end{eqnarray}
and compute the separability criterion $Q_0$  which yields the maximal value $\frac{1}{2}$ (independently of $n$), while $Q_m$ is always zero, i.e. does not detect GME. Whereas for a W-state
\begin{eqnarray}\label{W}
|W_{n}\rangle&=& \frac{1}{\sqrt{n}}\sum_{i=1}^{n}\bigotimes_{k\not=i} |0\rangle_k\otimes |1\rangle_i\nonumber\\
\textit{e.g.:}\quad |W_3\rangle&=&\frac{1}{\sqrt{3}}\left(|001\rangle+|010\rangle+|100\rangle\right)
\end{eqnarray}
the criterion $Q_{m=1}$ gives the maximal value $1$, but the criterion $Q_0$ fails to detect GME. In general, the criteria $Q_m$ assume their respective maximal values for the corresponding $n$-partite Dicke-state with $m$ excitations
\begin{eqnarray}\label{dicke}
|D_n^m\rangle&=& \frac{1}{\tiny{\sqrt{\left(\begin{array}{c}n\\m\end{array}\right)}}}\sum_{\{\beta\}}|d_\beta\rangle\;,
\end{eqnarray}
where the set of indices $\{\beta\}$ corresponds to the respective subsystems of excitations and the sum is taken over all inequivalent sets $\{\beta\}$ fulfilling $|\{\beta\}|=m$. In the case of $m=1$, these states are the $W$-states.

A characteristic trait of these GME detection criteria is their noninvariance under local unitary transformations, which stems from their formulation via density matrix elements (in contrast to other criteria, which are often based on e.g. eigenvalues). This has the advantage of dramatically reducing the required measurement complexity for their computation and thus experimental feasibility~\cite{HiesmayrGMENeutron}, but comes at a price: In order to detect GME in a given state, the basis in which the density matrix elements are computed might need to be adapted to the state in question. We will show, however, that this is surprisingly not necessary in a wide variety of cases.

\section{Device-Independent Test for GME}
Consider the following scenario. A source produces an unknown multipartite state, which is supposed to be GME. The individual parties receiving the particles want to verify the presence of GME without going through the trouble of performing a full quantum state tomography (i.e. by only performing a smaller number of measurements than needed for a state tomography).\\
Without complete knowledge of the exact state, this problem can in principal only be solved with a finite probability $p<1$, as there is no way to guarantee a choice of observables yielding a definite answer to the entanglement problem. As the GME detection inequalities $Q_i$ only require small numbers of density matrix elements for computation and have a quite high detection efficiency~\cite{hmgh,huber_dicke}, they offer a good approach to this problem.\\
By using the Haar measure \cite{spengler_haar}
\beq \label{eq_integral} p_{F} = \int \mathrm{dU_L} \ |J| \ (\biggl\lbrace\begin{array}{l}1\dots\textrm{if}\quad  F(U_L\cdot\rho\cdot U_L^\dagger)>0\\
0\dots\textrm{if}\quad  F(U_L\cdot\rho\cdot U_L^\dagger)\leq 0 \end{array}) \eeq
with, e.g., $F(\rho)=Q_i(\rho)$ we obtain the a-priori probability for the detection criterion $Q_i$ to detect $\rho$ to be GME in a randomly chosen basis, where the integral runs over all local unitary transformations $U_L$ (or, depending on the specific scenario, only a certain subgroup) and $|J|$ is the absolute value of the Jacobi-determinant, such that
\beq \int \mathrm{dU_L} \ |J| = 1 \quad .\eeq
These integrals can be computed numerically or, in some cases, even analytically (e.g. by means of the composite parametrisation from Ref.~\cite{spengler_haar}).

If necessary, the detection probability can be increased by combining several $Q_i$ in different measurement bases, i.e. choosing
\beq
F(\rho)=\max\left(Q_i(\rho), Q_i(U_x\cdot\rho\cdot U_x^\dagger), Q_j(\rho) , ...\right)
\eeq
where $U_x$ is e.g. the local unitary rotation to $\sigma_x$-eigenstates, i.e. the Hadamard matrix
\beq U_x = \left(\begin{array}{cc} \frac{1}{\sqrt{2}} & \frac{1}{\sqrt{2}} \\ \frac{1}{\sqrt{2}} & -\frac{1}{\sqrt{2}} \end{array}\right)^{\otimes n}\;. \eeq
Note that the scenario of an unknown state produced by the source is fully equivalent to a known state which is perturbed by the used quantum channels before it can reach the individual parties or the case in which the workings of the detectors are unknown. For sake of simplicity, in the following we focus for simplicity on the unknown state scenario.

\section{Examples}

Let us illustrate this approach by means of some simple examples, for which we consider a three-qubit-system. Depending on the actual knowledge over the state %(e.g. the source might in fact produce a known state, which is then somehow perturbed by the used quantum channels, before the individual particles reach their destinations),
the integration in eq. (\ref{eq_integral}) may cover different unitary groups. We start with the least favorable case when nothing is known about the local basis in which the state is prepared and proceed with the case when some knowledge about the symmetry is available.

\subsection{No Prior Knowledge: $U_\alpha \otimes U_\beta \otimes U_\gamma$}
If there is absolutely no prior knowledge available about the three local bases the state is prepared in, the associated unitary group is $\mathcal{SU}(2)^3$. Even in this least favorable case, the probability for detecting GME using a single, randomly guessed measurement basis is surprisingly high. If, for example, the produced state is a GHZ-state
\beq \rho_U = \ket{\Psi}\bra{\Psi} \quad \mathrm{with} \quad \ket{\Psi}=\frac{U_\alpha\otimes U_\beta\otimes U_\gamma}{\sqrt{2}}(\ket{000}+\ket{111}) \quad ,\eeq
GME can still be detected in more than $25 \%$ of all randomly chosen bases, using only a single separability criterion, namely $Q_1$. It is a surprising result here that not the criterion $Q_0$, the one designed to detect the GHZ-state in the computational basis, gives the maximal value. Using $Q_0$ one detects GME with $18\%$ probability.

\begin{figure}[ht!]\centering\includegraphics[width=0.55\textwidth]{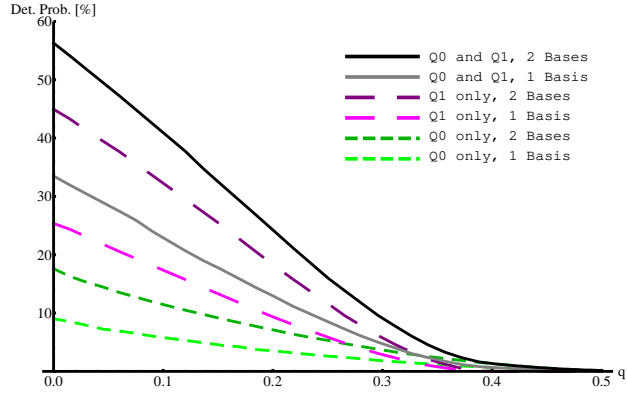}\caption{Probability of detecting GME in a GHZ-type isotropic state ($\rho = (1-q) \ket{GHZ}\bra{GHZ} +\frac{q}{8}\id$) in a random basis, by means of different separability criteria. The two short-dashed (green) lines correspond to using only $Q_0$, the long-dashed (purple) lines to $Q_1$ and the solid (gray/black) lines to both, where in each case the lower (lighter) line represents measurements in only one basis, while in the upper (darker) line two mutually orthogonal measurement bases are used. Note that for $q = 0.571$ this state becomes biseparable \cite{taming}.}\label{fig_ghz_abc}\end{figure}

If $Q_0$ and $Q_1$ are used, as well as a second measurement direction (orthogonal to the first), this ratio increases to more than $56 \%$. Note that this is the highest possible effort using the separability criteria $Q_i$, as for three qubits there are only two such criteria and only two independent and inequivalent measurement bases. Even if the produced state is not pure, but perturbed by some kind of noise, the detection probability only slowly decreases with the amount of the noise, as illustrated in Fig.~\ref{fig_ghz_abc}.

\begin{figure}[ht!]\centering\includegraphics[width=0.55\textwidth]{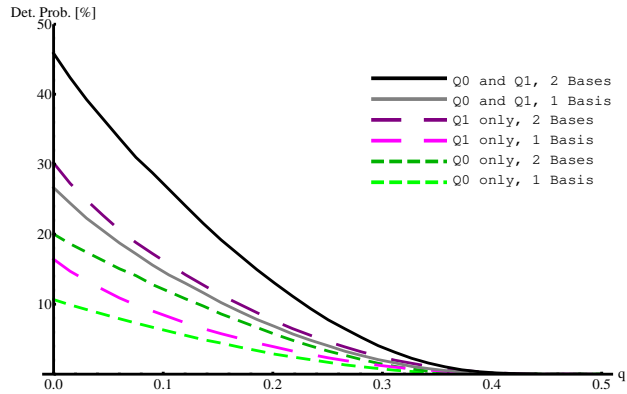}\caption{Probability of detecting GME in a W-type isotropic state ($\rho = (1-q) \ket{W}\bra{W} +\frac{q}{8}\id$) in a random basis, by means of different separability criteria. The two short-dashed (green) lines correspond to using only $Q_0$, the long-dashed (purple) lines to $Q_1$ and the solid (gray/black) lines to both, where in each case the lower (lighter) line represents measurements in only one basis, while in the upper (darker) line two mutually orthogonal measurement bases are used. Note that for $q = 0.521$ this state becomes biseparable \cite{taming}.}\label{fig_w_abc}\end{figure}

In a similar setup, $W$-states $(\ref{W})$ are detected with only slightly lower efficiency: the corresponding detection probabilities range from $10,7 \%$ in the worst case (using only $Q_0$ in one basis) to $45,9 \%$ (for the combination of $Q_0$ and $Q_1$ in two mutually orthogonal measurement bases each), as shown in Fig.~\ref{fig_w_abc}. Here we do not observe the interchanged roles of the quantities $Q_0$ and $Q_1$.

\subsection{Subsystem Symmetry: $U^{\otimes 3}$}
\begin{figure}[ht!]\centering\includegraphics[width=0.55\textwidth]{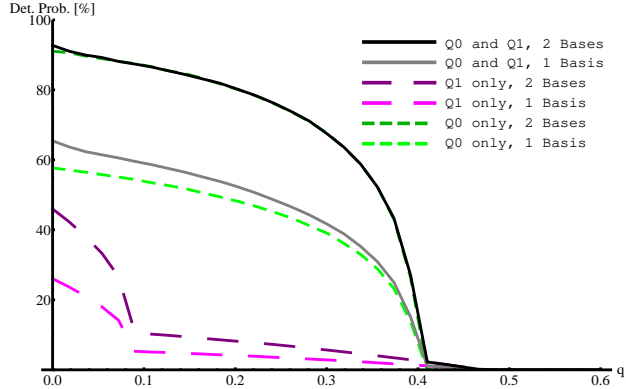}\caption{Probability of detecting GME in a W-type isotropic state ($\rho = (1-q) \ket{W}\bra{W} +\frac{q}{8}\id$) in a random symmetric basis, by means of different separability criteria. The two short-dashed (green) lines correspond to using only $Q_0$ (note that the dark green short-dashed line is mostly below the black one), the long-dashed (purple) lines to $Q_1$ and the solid (gray/black) lines to both, where in each case the lower (lighter) line represents measurements in only one basis, while in the upper (darker) line two mutually orthogonal measurement bases are used. Note that for $q = 0.521$ this state becomes biseparable \cite{taming}. The somewhat unusual behaviour of the curve stems from two disjoint areas of $U$, for which $Q_1$ is positive, one of which vanishes at $q \approx 0.1$.}\label{fig_w_aaa}\end{figure}

With the knowledge of the source's output state, the GME detection effectivity increases dramatically. If, e.g., the state is known to be symmetric under particle-exchange, only the symmetric subgroup of the group of all local-unitary transformations
\beq (\mathcal{SU}(2)^3)_{symm} \cong \mathcal{SU}(2) \eeq
has to be integrated over. In this case, the probability of GME detection increases to more than $90 \%$, if both criteria and two measurement bases are used (for both the GHZ- and the W-state). In fact, for the W-state, a detection probability of about $91 \%$ can be achieved even using only $Q_0$, as illustrated in Fig.~\ref{fig_w_aaa}.

In cases with high degree of symmetry we are even able to obtain analytical expressions for the probabilities in question. For example, for the GHZ-state and only the criterion $Q_0$ in a single basis, the detection probability computes to (in the following $p_i$ denotes the probability $p_F$ with $F=Q_i$)
\beq\label{analyticresult} p_0 =1+\frac{3 \sqrt{2} (2 K(-\frac{1}{8}) - 3 \Pi(-\frac{1}{2}, -\frac{1}{8}))}{ 2 \pi} \approx 0.52966\;    , \eeq
where $K(x) = \int_0^{\frac{\pi}{2}}\frac{d\theta}{\sqrt{1-x^2\sin^2(\theta)}}$ is the complete elliptic integral of the first kind, and $\Pi(x,y) = \int_0^\frac{\pi}{2}\frac{d\theta}{(1-x\sin^2(\theta))\sqrt{1-y^2\sin^2(\theta)}}$ is the complete elliptic integral of the third kind.

\section{Discussion}
It can be seen from the presented plots (Figs. \ref{fig_ghz_abc}, \ref{fig_w_abc} and \ref{fig_w_aaa}) that the two GME detection criteria ($Q_0$ and $Q_1$) in the tripartite case have very different detection behaviour. Although $Q_0$ and $Q_1$ are specifically designed to detect GHZ- and W-type states (respectively), this is not reflected in the detection probabilities. Instead, $Q_0$ seems much more suitable to detect GME in symmetric bases, while $Q_1$ is much more efficient for nonsymmetric bases. In this context, the two criteria complement each other quite well, such that the combined detection power is highly satisfactory (for both GHZ-type and W-type states).\\
While all above results were obtained by numerical methods, the integral (\ref{eq_integral}) can be solved analytically in many cases, even for general numbers $n$ of subsystems (as long as the number of parameters is not too high, i.e. if the degree of symmetry is sufficient). For example, for the $n$-partite W-state in a symmetric basis, the probability to be detected by $Q_1$ (using only one basis of measurement) computes to
\beq p_1(n) = \frac{1}{n}\left(1+\sqrt{\frac{n-1}{n-2}}-2\sqrt{\frac{n-1}{n}}\right) \eeq
in which case $Q_0$ yields
\beq p_0(n) = \left\lbrace\begin{array}{c} \frac{1}{\sqrt{3}} \approx 57,7 \%\quad\textrm{for}\quad n=3\\ 0 \quad\textrm{for}\quad n>3\end{array}\right.\;.\eeq
Analytic expressions for less symmetric scenarios can often be computed, but as can be recognized from Eq.~(\ref{analyticresult}) are much more cumbersome in general.\\
It is not surprising that for increasing $n$ the individual detection probabilities decrease, as also the number of different types of GME increases (along with the number of detection criteria $Q_i$). For example, for four qubits there is also the 2-Dicke state (see Eq.~(\ref{dicke}))
\beq \ket{D_4^2}=\frac{1}{\sqrt{6}}\left(\ket{0011}+\ket{0101}+\ket{0110}+\ket{1001}+\ket{1010}+\ket{1100}\right) \eeq
and the corresponding detection criterion $Q_2$, both of which can straightforwardly be utilised in our device-independent detection scheme. For the 4 qubit 2-Dicke state, the detection probabilities in the symmetric case read
\beq \begin{array}{c} p_0 = \frac{1}{\sqrt{3+\sqrt{6}}} \approx 42,8 \% \\
p_1 = \frac{1}{2}\left(\sqrt{2}-1\right) \approx 20,7 \% \\
p_2 = 1 - \frac{\sqrt{8-2\sqrt{3}}}{3} \approx 29,0 \% \\
p_{012} = \frac{1}{6}\left(3+3\sqrt{2}+2\sqrt{3(3-\sqrt{6})}-2\sqrt{4+\sqrt{13}}\right) \approx 71,6 \% \end{array}\eeq
where the $p_i$ are the individual detection probabilities using only $Q_i$, and $p_{012}$ is the detection probability of using $Q_0$, $Q_1$ and $Q_2$ simultaneously (in a single measurement basis).\\
An issue of great interest in the context of any device-independent characterisation scheme for quantum states is the measurement. The quantities $Q_i$ can be measured comparatively easily, e.g. by decomposing the density-matrix-elements into expectation values of Pauli operators (which is particularly efficient for the diagonal elements of the density matrix, as these can all be obtained using only a single measurement setting per used basis).

In particular, for a single neutron propagating through an interferometer the criteria $Q_0$ and $Q_1$ were able to prove
the generation of genuinely tripartite entangled W-like state families and a GHZ-like state with high fidelity~\cite{HiesmayrGMENeutron}.
A single neutron can be entangled in its path-degrees of freedom, its spin-degrees of freedom and its energy-degrees of freedom, thus constituting an effective three-qubit system. In this case, a state tomography is unfeasible for technical reasons (maybe even impossible with today's equipment), thus the criteria, $Q_0$ and $Q_1$, serve as good experimental test to verify GME. Even if noise was included the GME between outer and inner degrees of freedom could be witnessed. Our proposed device independent test predicts that the detection probabilities would also be high if one was not able to construct a measurement device in the optimal basis choice, but slightly rotated.

Let us also comment on secret sharing protocols. The main idea is to divide a secret into several shares and distribute these shares to $n$ parties such that the secret (or even a part of it) cannot be reconstructed by a subset of parties. Hence, only if all parties work together the secret can be revealed. This scheme was brought to quantum physics in 1999~\cite{HilleryBuzekBerthiaume} by exploring the GME of GHZ states. In Ref.~\cite{qss} an experimentally feasible security check was introduced by changing the original protocol and allowing the distributer to choose randomly between the GHZ and a rotated GHZ state. The point is that the unrotated GHZ state maximally violates $Q_0$ while the rotated GHZ state does not violate $Q_0$ (in the standard basis). Of course, there exists a $\tilde{Q}_0(\rho) = Q_0(U\cdot\rho\cdot U^\dagger)$ which is maximally violated for the rotated GHZ state while it is not violated for the un-rotated one, simply changing the bases. Since an eavesdropper or a dishonest party does not know which state, unrotated or rotated GHZ state, is distributed, there exists no strategy to reveal the secret without being detected (due to the complementary detection properties of the quantities  $Q_0$ and $\tilde{Q}_0$). Consequently, our proposed device independent test can be used to analyse the measurement device regarding its trustfulness probabilistically and, herewith, analyse the security of a given quantum cryptographic protocol.

\section{Conclusions}

We introduced a method of detecting GME (genuine multipartite entanglement) for states in unknown bases, i.e. device-independently. Alternatively, our method applies to the situation when one does not know how a channel decoheres or one has only partial information about the very working of a detector. Device-independence is achieved by applying certain separability criteria to the (partially unknown) state, which yields decisive results with comparatively high probabilities (which can be increased further by combining several separability criteria and/or several mutually unbiased measurement bases). These probabilities are spread over a wide range of values, depending not only on the actual state but also on the knowledge present about them. Typical values for these probabilities for highly GME states range from several per cent to several tens of per cent.

A rather counter-intuitive result of our work concerns the used GME-criteria $Q_i$ themselves: While $Q_0$ and $Q_1$ were designed to detect GME in GHZ- and W-states, respectively, this is not reflected in their a-priori GME-detection probabilities. Instead, $Q_0$ seems more suitable to detect GME in symmetric states (e.g. symmetric bases for either GHZ- or W-states), while $Q_1$ is more efficient in detecting GME in asymmetric states (such as GHZ- or W-states in random bases).

A possible application of our method are situations where the optimal measurement settings are not available (e.g. due to experimental reasons). In this case, our method allows for an estimation on the probability of success for the used setting.

We also discussed how the used criteria can be experimentally implemented using fewest possible measurement settings, thus minimising the experimental complexity. In this way, our scheme can also be understood as an intermediate way to detect GME: the probability of success increases steadily with the number of used measurements (i.e. the number of used criteria and bases), such that it interpolates between common GME detection criteria and a full quantum state tomography.

Our method may also be applicable to other scenarios, such as analysing the security of quantum cryptography and secret sharing protocols.

\section{Acknowledgements}
The authors would like to thank Christoph Spengler for his source code for parameterising unitary groups and the Haar measure. AG gratefully acknowledges the Austrian Research Fund project (FWF-P21947N16). {\L}R acknowledges the Polish Ministry of Science and Higher Education project No. IP2011 046871. BCH acknowledges gratefully the Austrian Science Fund (FWF-P23627-N16).

\end{document}